# Proposal for a multilevel university cybermetric analysis model

**Enrique Orduña-Malea**[1*] **and José-Antonio Ontalba-Ruipérez**[1]

[1]Department of Audiovisual Communication, Documentation, and History of Art. Polytechnic University of Valencia (UPV), Valencia, Spain.
Camino de Vera s/n, Valencia 46022, Spain.
* e-mail: enorma@upv.es

**Abstract** Universities' online seats have gradually become complex systems of dynamic information where all their institutions and services are linked and potentially accessible. These online seats now constitute a central node around which universities construct and document their main activities and services.
This information can be quantitative measured by cybermetric techniques in order to design university web rankings, taking the university as a global reference unit. However, previous research into web subunits shows that it is possible to carry out systemic web analyses, which open up the possibility of carrying out studies which address university diversity, necessary for both describing the university in greater detail and for establishing comparable ranking units.
To address this issue, a multilevel university cybermetric analysis model is proposed, based on parts (core and satellite), levels (institutional and external) and sublevels (contour and internal), providing a deeper analysis of institutions. Finally the model is integrated into another which is independent of the technique used, and applied by analysing Harvard University as an example of use.

## 1. Introduction

From the outset, the Internet and the World Wide Web have been very closely linked to universities, both in their design, development and expansion, and in the impact of these services on the way the institution works (Goldfarb 2006; Castells 2001; Berners-Lee and Fischetti 2000).

The accessibility, affordability and personalisation of these information and communication technologies (ICT) are factors in their penetration and impact on all walks of life; higher education institutions are particularly influenced by four disruptive forces (Katz 2008a):
  - Unbundling. Content producers and creators can deconstruct services, allowing greater flexibility in services on offer.
  - Demand-Pull. Users can assemble content to suit their own interests.
  - Ubiquitous access. Online content is accessible from a multitude of devices, without geographic or temporal restrictions.
  - Intellectual property. There is a need for real and effective rules that describe how individuals enter and exit the online university community.

Since universities are organisations that depend on attracting, developing and organising human talent for the purpose of creating and disseminating intellectual capital (Katz 2008a), some of the activities they should consider with respect to the impact of Internet use are (Katz 2008b):
  - Develop new points of view on research processes, educational content, publications, software, and informational instruments and resources.
  - Find out how new communication tools (blogs, wikis, mashups, and many more) will interact with traditional academic approaches and thinking.

- Understand what data and information the organisation is responsible for, and how they will be stored and protected.
- Understand what influence the institution's digital resources can have on its reputation.
- Maintain and enhance the online identity of the institution.

## 1.1. Universities online

The Internet (as a protocol for communication between computers) and the World Wide Web (as a service containing multiple online services) extend the reach of universities, and they need to look at how to manage this expansion as well as the various ways in which the property of the institution may be developed, disseminated, commercialised and utilised online.

Therefore the creation, design and maintenance of an academic web seat must have clear objectives. To this end, Middleton, McConnell and Davidson (1999) propose a series of questions and answers that should be taken into account when designing a university web seat, which include the following (table 1):

Table 1. Objectives of an academic website

| QUESTION | ANSWER |
|---|---|
| Why have a website? | Means of communication between individuals and groups |
| | Means of accessing online facilities and services |
| | Tool for representing and promoting the institution |
| Whom does the website serve? | Internal users (captive market): staff and students |
| | External users (target market): alumni and prospective students, academics, business people, the media, etc. |
| What kind of information is required? | Promotional; value-added; and useful |

## 1.2. The complexity of the academic web seat

At the end of the 20th Century, the majority of university web seats were relatively small, the information they provided was homogeneous and they did not contribute much in terms of added value. The current situation, however, is very different: the online seat now constitutes the central node around which universities construct their online presence. The rest of the online services (e.g. email, file transfer, news services, etc.) are centralised on these websites which also serve as a document repository for the institution's activities (teaching, research, transfer) as well as the administrative, governing and management services (vice-rectors' offices, libraries, student services, etc.).

In this way, university online seats have gradually become complex systems of dynamic information where all the university institutions and services are linked and potentially accessible from a general URL, via subdomains and subdirectories. In addition, each of these may be associated to a different (or several) content management system (CMS) which independently manages the contents linked to it (i.e. a service, such as the library; a product, such as a repository; or an institution, such as a faculty), although the users are not aware of this when they are browsing the university website.

The complexity of managing and maintaining this kind of seat gave rise to the foundation, in 2006, of the HighEDWeb Professionals Association, the result of a merger between an alliance of Web professionals in New York (originally called HighEDWeb), and WebDevShare (an international conference for Web developers specializing in higher education).

This complexity derives fundamentally from the diversity of university functions and the heterogeneous audience. The third mission (transfer) adds to that complexity and diversity: the universities organise external events, manage university hospitals, are in charge of museums, radio and television stations, run sports teams, publication services, alumni services, manage patents – the list goes on – all of which have a greater or lesser online presence through the university seat and other external platforms (Aguillo 2009).

The complexity of the universities is thus transferred to the online seats that represent (and reflect) them. Saorín (2012) identifies, at present, 5 levels of content on a university online seat:
- Information spaces: correspond to the association of an online seat and a URL.
- Information products: both communication channels for institutions, organisations, or associations (e.g. departments), and specific products (e.g. an institutional repository).
- Tools/services: support applications, such as email or electronic administration.
- Contents: published units of information, associated to a URI.
- Digital objects: associated to a format and file.

1.2.1. Products and tools/services

Examples of products (which store digital contents and objects, and which are identified in specific information spaces through subdomains or subdirectories) are listed below, organised according to the university mission to which they correspond[1].

*Teaching activities*

These can be divided into, on the one hand, platforms geared towards the publication and dissemination of teaching materials (OpenCourseWare or virtual campuses) and, on the other hand, the online presence of institutional units focused on teaching, which is essentially what university departments are.

*Scientific activities*

With respect to the online platforms partially or totally dedicated to the research mission of the universities, those that stand out are institutional repositories, magazine platforms and university presses; with respect to research institutions with an online presence, those that stand out are research groups, institutes and research centres.

*Transfer activities*

Of particular interest are the online platforms of the "Research results transfer offices" (such as OTRI offices, located in the Spanish academic system) and the postgraduate educational centres, which are part of the lifelong learning framework. University hospitals are also in this category.

*Administrative activities*

Include areas of governance (such as vice-rectors' offices) and centres such as Schools, Colleges, and University Faculties. The information presented on these online seats is usually informative, administrative and corporate. Another kind is the personal profile pages of the teaching and research staff.

*Service activities*

Include alumni associations, sport services, diffusion of news and cultural activities, institutional radio or television channels, digital archives or collections, or university libraries. With respect to products, these may include blogs and video platforms, among others.

Table II presents examples of each of the aforementioned products and services:

**Table II. Examples of products, services and tools on a university online space**

| ACTIVITY | TYPE OF ENTITY | UNIVERSITY | URL |
|---|---|---|---|
| TEACHING | OCW Platform | MIT | ocw.mit.edu |
| | Virtual campus | Stanford University | bb.stanford.edu |
| | Department | Oxford University | economics.ox.ac.uk |
| RESEARCH | Institutional repository | National University of Singapore | scholarbank.nus.edu.sg |
| | Research group | University of Wolverhampton | cybermetrics.wlv.ac.uk |
| | Research institute | University of Michigan | umtri.umich.edu |
| | Research centre | University of Malaya | cenar.um.edu.my |
| TRANSFER | Research results transfer office (OTRI) | University of Seville | otri.us.es |
| | Learning centre | University of Delaware | lifelonglearning.udel.edu |
| | University hospital | Columbia University | cumc.columbia.edu |
| ADMINISTRATION | Vice-rectorates | University of Granada | investigacion.ugr.es |
| | Schools | University of Cambridge | trin.cam.ac.uk |
| | Colleges | University of California-Berkeley | ischool.berkeley.edu |
| | Faculty | University of Oslo | jus.uio.no |
| | Personal website | University of Chile | dcc.uchile.cl/~rbaeza |
| SERVICES | Alumni services | Cornell University | alumni.cornell.edu |
| | Sports services | University of Canada | athletics.utoronto.ca |
| | News services | Leiden University | news.leiden.edu |
| | Media services | University of Miami | umtv.miami.edu |
| | Digital archives and collections | TexasA&M University | chinaarchive.tamu.edu |
| | Libraries | University of California-Los Angeles | library.ucla.edu |
| | Blog platforms | Penn State University | blogs.psu.edu |

| | | |
|---|---|---|
| Video platforms | Polytechnic University of Valencia | politube.upv.es |
| Journal platform | Complutense University of Madrid | revistas.ucm.es |
| University press | University of Princeton | press.princeton.edu |

In a complementary way, universities have begun wide-scale use of resource-sharing tools (tool/service level, according to Saorín). These include news aggregators (like RSS or Atom), forums, blogs and microblogs, podcasts, chats, etc., which complement services offered by the aforementioned platforms.

Universities also create institutional accounts on websites that share videos (Youtube[3], Vimeo[4]), photographs (Flickr[5]), and presentations (Slideshare[6]), as well as on online social networks, both general (Facebook[7]) and academic (Academia.edu[8]). In these cases, the universities are expanding their presence onto platforms that are external to their academic web domains, and which in this paper will be termed "satellite".

1.2.2 The university's digital footprint

These institutions have so far published millions of pages, with rich, varied and in some cases, highly value-added, contents (Aguillo et al. 2008).

All the types of online seats thus far described (as well as others in existence) are created independently (universities no longer have a centralised control over their contents), but they are hosted by the university online seat, occasionally reflecting the existing hierarchical relations (see URLs in table II), constituting a complex system of documents which generates a clear digital footprint that aids in the creation and profiling of a clearly-defined institutional online identity (Tiscar 2009).

The scope and variety of this digital footprint is such that analysing it, both quantitatively and qualitatively, could bring to light information unobtainable through other research methods (Aguillo 2009). In this context, the discipline of cybermetrics, in its broader definition, provides the tools and the methodology necessary for a quantitative analysis of the information contained on university servers and seats; this aspect is set out below.

**1.3. Cybermetric analysis of the university**

University online seats have been employed as units of analysis in numerous cybermetric studies, which can be broadly divided into the following two nonexclusive groups:
- Studies focusing on the academic discipline itself, not the university (Thelwall et al. 2005).
- Studies describing and analysing university systems, and the relationships between them.

The latter category notably includes the design and compilation of university rankings, which are described below.

1.3.1. Ranking web of universities

The scale and scope of academic data gleaned from cybermetric techniques has enabled university web rankings to be compiled; Spain is particularly prominent in this field.

Buenadicha et al. (2001) undertook a pioneering project analysing Spanish universities, the aim being to develop a web seat evaluation index with which to compare Internet use by universities. The project resulted in the design for a Spanish university web ranking, sorted by 4 main categories: content, web page load time, accessibility and browsability.

Another project of note is that carried out by the Observatory for Audiovisual Contents[9] at the University of Salamanca, who compiled a ranking of Spanish universities based on the quality of their web pages (Acosta 2009)[10].

In 2004, the most important (that with the greatest impact) global university web ranking, from the Cybermetrics Lab, was initiated in Spain within the Spanish National Research Council (CSIC). The Ranking web of world universities proposes to improve the web impact factor (Ingwersen 1998) by means of the Webometric Ranking (WR)[11].

Despite natural corrections being made over the years, WR has proven that the results obtained are roughly similar to those obtained through other ranking systems with a very different methodology (Aguillo et al. 2006), which clearly indicates that online data, with adequate treatment and analysis, can be instrumental in identifying existing phenomena as well as reliably reflecting the universities' activities (figure 1).

**Figure. 1. Ranking Web of World Universities (source: webometrics.info)**

Other university rankings based on web indicators are the Web Popularity Ranking[12], the Ranking Universitario de transparencia Web[13] (University Web Transparency Ranking), Ranking Colleges using Google and OSS[14], and the original G-Factor[15] proposal. Another initiative is The Google College Rankings[16], which aimed to use the search engine to rank universities, but which seems to have now been abandoned.

**1.4. Analysis of cybermetric subunits**

Web analyses of university systems have occasionally centered on specific internal units (essentially departments); this has not yet occurred in the area of university web rankings. Although it has a greater degree of granularity, this type of analysis requires the units of measurement to be clearly established.

The most relevant studies performed on unit level are detailed below, together with a discussion on the identification and description of measurement units.

**1.4.1. Online university subunits**

The analysis of academic online seats has also focused on specific units, such as professors' personal pages (Thelwall and Harries 2004a; 2004b; Barjak et al. 2007), or specific content types, such as news articles (Yolku 2001).

The most studied subunit, however, is that which contains university departments and schools (Thelwall 2002a; 2003), although the results should be interpreted with caution, given the low number of links per department. Some of these studies have sought to establish a possible correlation between the productivity of the department and links received. For example, Thomas and Willet (2000) have studied the departments of librarianship and information science, although the study did not find a significant correlation between links and research performance. The same occurred in the study carried out by Chu, He and Thelwall (2003), who also encountered disparities between link metrics and the U.S. News rankings, in this case for information science schools.

Tang and Thelwall (2003) also show the low level of interlinking between history departments in the United States, but they discover that there are significant differences in the link patterns depending on the scientific area that is analysed. In the same way, important correlations between links and scientific output are detected in British departments of computer science (Li et al. 2003), and psychology and chemistry (Tang and Thelwall 2004).

Other interesting studies in this field are those carried out by Ortega and Aguillo (2007), as well as the doctoral thesis of Li (2005), which specifically focuses on the interlinking patterns between university departments.

1.4.2. Cybermetric units of measurement

As in any metrics-based discipline, it is necessary to establish what the most suitable units of measurement are, a complicated matter in cybermetrics given the particular properties of the electronic documents to be analysed; these have been subject to research by, among others, Thelwall (2002b).

The main problem is that the psychical unit and the unit of content do not match up exactly. An electronic file (which may or may not be a web file) can correspond to a book (or chapter of a book, or one or several selected pages), an article, image, application, directory, institutional pamphlet, etc. Counting a file can mean counting contents that are intellectually very different.

This issue is analysed by Thelwall (2002b), who proposes a conceptual analysis of different units of study that are problematic in terms of their definition and consideration. The starting point is the fact that there are diverse technical difficulties in identifying what it is that we should understand by web page. From a purely physical point of view, the "web page" is understood to be a unique HMTL-based file to which other files (web and non-web) can be associated, which can be independent or grouped together with other pages (web and non-web), and which is accessible online through an URL (Thelwall 2009).

This definition creates a series of problems for Thelwall:
- File format: should we only consider HTML (XML) files as web pages, or should we extend the definition to include any format that can be viewed with a web browser?
- Access mechanism: should only sites accessible via port 80, or encoded in HTTP, be considered, or any mechanism accessible through web browsers, such as FTP?

- Scope: should only public documents be considered, or should those that are private (on Intranets or Extranets) or that cannot be located (invisible Web) also be included?
- Permanence: should only static pages be considered, or should dynamic documents generated from databases also be considered?
- Number of files: should web pages composed of various file types (HTML, CSS, DTD, associated images, etc.) be counted as one single page, or one page per file?

Finally, the following definition of a web document is given (Thelwall 2009):

"A web document is a body of work with a consistent identifiable theme produced by a single author or collaborating team. It may consist of any number of part or whole unrestricted access electronic files retrievable over the Web using a modern browser".

This definition is characterised by the following elements: work, theme or genre, author (physical or institutional), files that constitute the work, and mode of access.

Consideration of the "genre" could be a solution to the problem. Counting pages (or collections of pages) by genre would mean counting the number of online instances, which would avoid the aforementioned problem of nonequivalence of content and format. However, the identification of these genres is considered to be very complicated (Crowston and Williams 2000).

An alternative to the study of genres is the alternative document model (ADM) proposed by Thelwall (2002b), who establishes 4 document-based models that are exclusively of Web organisation:
- Individual Web page: each HTML file is treated as a document for the purpose of extracting links.
- Directory: all HTML files in the same directory are treated as a document. All the links from all the files in the directory are combined and duplicates eliminated.
- Domain name: all HTML files with the same domain name are treated as a document.
- University: all pages belonging to a university are treated as a single document.

The ADM "university" is later called "Site" (Thelwall 2009), now characterised by the fact that it allows multiple domain names for the same unit of study, identified by its TLD. This concept of "Site", however, poses a conceptual problem by integrating "place" and "work". For this reason, Aguillo (1998) and Pareja et al. (2005) separate and distinguish the concepts of web page (physical dimension), web site (spatial dimension) and web seat (conceptual dimension):
- Web page: electronic file or set of files that constitute a document in HTML, i.e. hypertextual and multimedia, identifiable online with a unique URL.
- Web site: physical space connected to the Internet where the information is stored in electronic form, accessible via HTTP, i.e. a computer which acts as a web server with a unique IP address.
- Web seat: set of web pages linked hierarchically to a main page, representable through its URL, making up a documental unit, distinguishable from others, and an institutional unit, in which it is possible to identify authorship.
- For their part, Björneborn and Ingwersen (2004) propose an analysis based on different levels of granularity: micro level (pages, directories, small subseats), meso level (seats and large subseats) and macro level (TLDs or large second-level domains).

The distinction between site and seat is not always taken into account, and they are often defined with inappropriate or confusing names. For this reason, Ayan, Li and Kolak (2002) propose differentiating between logical domain (group of pages that are semantically and structurally related; the equivalent to an online seat) and physical domain (which is identified only by domain name; this corresponds to an online site).

## 2. Objectives

The preceding sections have described the current complexity of the academic website, and its constitution as unit of cybermetric analysis, both in studies of university systems and in creating university rankings.

All the university web ranking initiatives seen in section 1.3 use the university as a whole as a unit of analysis; however, research into subunits shows that it is possible to carry out systemic analyses of universities following cybermetric techniques which, given the quantity and variety of documentation stored on subdomains and subdirectories (widely covered in section 1.2), open up the possibility of carrying out studies which address university diversity (Van Vught 2009), necessary for both describing the university in greater detail (systemic intrauniversity analysis) and for establishing comparable ranking units (systemic interuniversity analysis).

These systemic studies require the correct identification and definition of the units of measurement, and the establishment of a conceptual model (in order to facilitate the acquisition of information structured on various levels), which constitute the main objectives of this study.

Therefore the specific objectives are the following:
- To identify and propose a new definition of the different units of measurement employed in cybermetric analysis, extending the concept of web to that of "online".
- To propose, through the establishment of units of measurement, a multilevel university cybermetric analysis model, which allows structured gathering of information on online universities, and which will serve as the basis of future university ranking systems.
- To integrate this model into another which is wider, more conceptual and independent of the technique used, that will permit future comparison with other analyses produced with the same general model, but with a different method of analysis.
- To apply the proposed model by analysing the top university in the Ranking Web of Universities ranking (2012 edition): Harvard University.

## 3. Methodology

The development of this model is rooted in the theoretical study of cybermetric university analysis units. It first identifies the concept of web document and subsequently proposes a classification of logical units necessary in establishing the model of analysis.

Following this, the conceptual model of systemic multilevel university analysis proposed by Orduña-Malea (2012) is used as a reference model that is independent of the technique employed and which takes into account the attributes of university diversity and multidimensionality. The model is adapted to be applied through cybermetric techniques.

The internal and external levels are demonstrated through entities belonging to the Complutense University of Madrid (the highest ranked Spanish university on the Ranking Web), due to this institution clearly shows the diversity of activities and a wider casuistry of URL syntaxes, necessary to assess both the need for internal analysis and its complexity (both technical and organizational). The data were manually collated in January 2011, and one URL -from each found casuistry- was randomly selected, to illustrate the diversity of URL creation process.

The full application of the model is performed by analysing Harvard University (the highest ranked university on the Ranking Web). For that purpose both core and satellite URLs are collected, and measured at institutional and external levels (levels and other elements of the model will be defined and explained in section 4.2):
- The core contour level is composed by "harvard.edu" official URL.
- The core internal level is composed only by third-level sub-domains ("x.harvard.edu/x") in order to facilitate the test analysis. In any case, the measurement of third-level sub-domains implies the measurement of the remaining levels, and considering that subdirectories have methodological problems to be accurate analysed, the selected sample is considered representative. All URLs were gathered by browsing the official website, along May, 2012.
- The satellite levels (both contour and internal) were identified through the Harvard Social service. The platforms considered are the following: Academia, Facebook, Flickr, Twitter, and Youtube. The selection of these platforms is exploratory and only meant to show the webometric performance of some Harvard University channels on some social applications.
- Institutional measures (both core and satellite) were measured by count page indicator, using <site:> command on Google.
- External measures (both core and satellite) were measure by URL mention indicator, using: <"URL" –site:harvard.edu> command on Google for core, and <"URL" –site:domain.tld> for satellite (substituting "domain.tld" for each platform: academia.edu, twitter.com, etc.).

## 4. Results

### 4.1. Proposal and definition of units of cybermetric analysis

In light of the problems previously detected, the concept of "website" is expanded to "online seat": a unit of measurement constituted by files that do not necessarily have to be web files, as is often the case (office computing files, graphics, multimedia, etc., are taken into account in webometric analyses, and also generate and receive hyperlinks).

Therefore, all the aforementioned considerations confirm the need to replace the attribute "web" with that of electronic file or DLO (Document Like Object)[17] accessible online, more in line with the general interest subject of cybermetrics.

The following formal considerations are also proposed:

*Online site*

An "online site" implies a specifically delimited online space, defined by its physical size (located on a proprietary or contracted server) and its name or identity (online domain).

The name of the "online domain" is composed of two elements or levels:
- Top-level domain name (generic[18] or geographic[19]). Only functions as an attribute, delimiting the "name type" (examples: .com, .org, .es, .fr, etc.).
- Second-level domain name. Functions as identifier of the product, service or entity (physical or legal) of the corresponding online space (examples: upv.es, nike.com).

The online site does not imply that the information is in web format, and the online domain only implies the formal identification of an available online space.

The online site is composed of one or more electronic files, which are public or private access depending on the restrictions imposed by the administrator of the online domain and the host.

The files located on this online space can be grouped logically in subdomains and subdirectories. Each of these groups is considered to be an online subsite.

Each file, subsite and online site is accessible and can be located online through a URI (URL or URN)[20].

*Online seat*

When the different files that make up a site (or one or various subsites within the same site) constitute a formally distinct documental unit (mainly due to authorship or thematic content), then this site or subsite will be called online seat.

**4.2. Multilevel cybermetric analysis model**

From the preceding section it can be inferred that the unit of measurement in cybermetric analysis (applied in this case to universities) must necessarily be defined both by a seat (logical, identifying unit) and by a site (physical, delimiting unit).

Consequently, and taking the Orduña-Malea model as a reference, the following elements of cybermetric measurement are proposed: core and satellite. Each of them can be divided into institutional and external, and applied at contour or internal sublevels.

These levels are summarized in detail in table III, where are displayed the separate levels, sublevels and unit types:

Table III. Multilevel analysis model: levels, sublevels and units

| SUPER LEVEL | LEVEL | DESCRIPTION |
|---|---|---|
| CORE | INSTITUTIONAL | Files created **within** the university website/unit. |
|  | EXTERNAL | Files created **outside** the university website/unit which mention the university/unit. |
| SATELLITE | INSTITUTIONAL | Files created on an **external platform** (Youtube, Twitter, etc.), by a university/unit. |
|  | EXTERNAL | Files created **outside** the university/unit satellite, which mention the satellite. |
|  | **SUBLEVEL** | **DESCRIPTION** |
|  | CONTOUR | The entire **University** is considered. |
|  | INTERNAL | A university **unit** is considered. |
|  | **UNITS** | **DESCRIPTION** |
|  | INSTITUTIONS | Websites of institutional university **entities**. |
|  | PRODUCTS | Websites of university **services and platforms**. |

4.2.1. Core level: institutional (direct measurements)

Core institutional measurements are related to the creation of content by the institution.

1. *Contour sublevel (general measurements)*

These refer to the analysis of the entire university as an institution, without taking into account the different university missions or the different entities that constitute it.

From a cybermetric approach, these will apply to all documentation published and deposited in the online academic domain of the institution. In this sense, it is represented totally and completely by the URL of the university, therefore all the indicators whose measurements are based on an analysis of the university's general academic domain will be considered contour measurements (for example, "ucm.es" and "harvard.edu").

2. *Internal sublevel (functional measurements)*

These focus on the activity of different and distinct university entities. In addition, these entities may be associated with the various functions or missions of the university. For example, the departments can be related to teaching activity and the research centres to scientific activity. These units can correspond to institutions or products.

From a cybermetric approach, internal measurements will be considered to be all those that measure web documentation published in a specific subdomain or subdirectory of the upper general domain assigned to the university, which in turn is associated to a specific university entity or service, e.g. "mat.ucm.es".

The diversity of these units (both in their functions and the manner of identifying them through their URL) and the sheer number of them, make this level the most difficult to identify and evaluate.

Table IV shows examples of internal units for different types of entities[21]. In each category of entity different configurations can be seen in the way the URL syntax is generated.

**Table IV. Internal level of cybermetric analysis by university unit**

| UNIT BY LEVEL | URL |
|---|---|
| **SCHOOLS** | |
| Nursing, Physiotherapy and Chiropody | ucm.es/centros/webs/euenfer/ |
| **FACULTIES** | |
| Fine Arts | ucm.es/centros/webs/fbartes/ |
| Mathematical Sciences | mat.ucm.es/ |
| Chemical Sciences | ucm.es/info/ccquim/ |
| Philosophy | fs-morente.filos.ucm.es/ |
| Computer Science | fdi.ucm.es/ |
| **DEPARTMENTS** | |
| Algebra | mat.ucm.es/deptos/al/ |
| Regional Geographical Analysis and Physical Geography | ucm.es/info/agrygfdp/Web/ |
| Anatomy and Human Embryology I | ucm.es/centros/webs/d529/ portal.ucm.es/web/anatomiai/ |
| Social Anthropology | ucm.es/info/dptoants/ |
| Cell Biology 3 | biocel.bio.ucm.es/ |
| Materials Physics | material.fis.ucm.es/ |
| Personality, Assessment and Psychological Treatment II | forteza.sis.ucm.es/dpto/ |
| **RESEARCH INSTITUTES** | |
| Economic Analysis (ICAE) | ucm.es/icae ucm.es/info/icae/ |
| Industrial and Financial Analysis | ucm.es/BUCM/cee/iaif/ |
| Biofunctional Studies | ieb.ucm.es/ |
| Ramón Castroviejo Ophthalmologic Research | ucm.es/info/iiorc/ |
| Interdisciplinary Mathematics (IMI) | mat.ucm.es/imi/ |
| Mediation and Conflict Management | ucm.es/centros/webs/iu5022/ |
| **RESEARCH GROUPS** | |
| Giftedness and Talent (ACYT) | ucm.es/info/sees/ |
| Quantitative analysis of economic policy and financial markets | ucm.es/info/ecocuan/anc/grupo/ |
| Architecture and technology of computing systems | artecs.dacya.ucm.es/ |
| Space astronomy – optimal astronomic resources management | mat.ucm.es/wso/ |
| Bioclimatology and biogeography | ucm.es/info/enviroveg/ |
| Asymptotic behaviour and dynamics of differential equations | mat.ucm.es/~cadedif/ |
| Veterinary control of microorganisms | ucm.es/centros/webs/gi5080/ |
| Formal analysis and design of software systems (FADOSS) | maude.sip.ucm.es/fadoss/ |
| Plant evolutionary ecology and ecological restoration | linneo.bio.ucm.es/balaguer/EvoEco/ |
| Proteins | bbm1.ucm.es/public_html/res/prot/ |
| Reproductive physiology of lagomorphs | ucm.es/info/fisani/sigue/ |
| Functional genomics of yeast and fungi | ucm.es/info/mfar/U4/ |
| Analysis, security and systems (GASS) | gass.ucm.es/ |

It can be observed that there is neither a pattern nor a specific policy in DNS management. Both subdirectories and subdomains are used (or both), as well as different hierarchical levels. For example, the Nursing School uses 3 subdirectories, the Fine Arts Faculty one subdomain, the Department of Materials Physics two subdomains, or the FADOSS Research Group two subdomains and one subdirectory, amongst other configurations.

On the other hand, there is no observable pattern in the classification. The subdirectories "centros" and "info" are used in all unit types. The subdirectory "dept" is not used in all departments (the variant "depts" is even used at times), and when it is used, it occasionally signifies the last level or may even contain a lower level subdirectory.

Finally, there are URLs that redirect to other valid internal URLs (i.e. the Institute of Economic Analysis), others that redirect to URLs external to the university (i.e. the Ramón Castroviejo Research Institute), or other URLs not associated to the online seat, since they are not hierarchically integrated into all the pages of the institution (Queueing Theory Group).

4.2.2. Core level: external (reputational measurements)

The core external measurements relate to the presence of the university (or specific units) in places external to the institution.

From a cybermetrics approach, these measurements are analogously linked to the presence of universities in locations external to their web domain, that is, to citation measurements (e.g. links) and audience.

Thus the number of times that a particular university (or unit) is named (invocation or citation), linked or visited online constitutes examples of core external measurements.

In turn, this level can be broken down into a contour sublevel and an entity sublevel according to whether the analysis is of the university in its entirety or a specific unit or service (table V).

Table V. External measurement level. Entities, units and possible measurements

| | |
|---|---|
| **EXTERNAL LEVEL. CONTOUR SUBLEVEL** | |
| **Entity** | Complutense University of Madrid |
| **Unit** | "ucm.es" |
| **Measurement** | |
| Hypertextual citation | linkdomain:ucm.es –site:ucm.es |
| Textual citation | "universidad complutense de Madrid" –site:ucm.es |
| **EXTERNAL LEVEL. INTERNAL SUBLEVEL** | |
| **Entity** | Department of Librarianship Information Science |
| **Unit** | "ucm.es/centros/webs/d168" |
| **Measurement** | |
| Hypertextual citation | linkdomain:http://www.ucm.es/centros/webs/d168 –site:ucm.es |
| Textual citation | "Departamento de Biblioteconomía y Documentación de la Universidad Complutense de Madrid" –site:ucm.es |

Both institutional and external levels can be merged into "core level", in opposition to satellite level, describe below.

4.2.3. Satellite level (extension measurements)

Finally, there is a second general level of online analysis (not included in the original model), made up of subdomains and subdirectories that belong to a university (or to any of its units), but are located outside its general domain, specifically in domains that belong to content-sharing platforms (e.g. Youtube, Academia, Twitter, Facebook, etc.). In this study, these are classified as "satellite" elements. In this sense, the satellite level measurements are completely analogous to core measurements, the difference being that they are applied to domains that are external to the university (table VI).

**Table VI. Satellite measurement level. Entities, units and possible measurements**

| | |
|---|---|
| **SATELLITE LEVEL. INSTITUTIONAL SUBLEVEL** | |
| **CONTOUR** | |
| Entity | Complutense University of Madrid |
| Unit | "ucm.academia.edu" (satellite in Academia.edu network) |
| **INTERNAL** | |
| Entity | Department of Librarianship and Information Science |
| Unit | "ucm.academia.edu/Departments/Biblioteconomía_y_Documentación" |
| Measurement (size) | "site:http://ucm.academia.edu/Departments/Biblioteconomía_y_Documentación –site:academia.edu" |
| **SATELLITE LEVEL. EXTERNAL SUBLEVEL** | |
| **CONTOUR** | |
| Entity | Complutense University of Madrid |
| Unit | "ucm.academia.edu" |
| Measurement (hypertextual citation) | "linkdomain:ucm.academia.edu –site:academia.edu" |
| **INTERNAL** | |
| Entity | Department of Librarianship and Information Science |
| Unit | "ucm.academia.edu/Departments/Biblioteconomía_y_Documentación" |
| Measurement (hypertextual citation) | "linkdomain: ucm.academia.edu/Departments/Biblioteconomía_y_Documentación –site:academia.edu" |

Moreover, in some cases, some unit level elements can be separated out within these satellites (upper level subdirectories or subdomains associated with particular university institutions or services). For example, a department can have a channel on Youtube regardless of whether the university also has one.

Finally, figure 2 shows a graphical representation of the Harvard University URL, in which each sphere represents the web space of a domain, subdomain or directory. The external level is represented by three possible web-spaces which can potentially link any part (core or satellite), level (institutional or external) or sublevel (contour and internal).

**Figure 2. Multilevel analysis model. Example for Harvard University**

### 4.3. Harvard University: a case study

The analysis of Harvard University website is divided into core and satellite levels, which are described below.

### 4.3.1. Core level at Harvard University

Following the proposed model, core level is formed by 1 contour level URL ("harvard.edu"), and 187 internal URLs, all of them third-level subdomains. The remaining subdomain levels and subdirectories have been avoided as indicated in method section, with the purpose of simplifying the example.

Due to its extension, the complete list of URLs and their corresponding institutions is free available for perusal in an annexed spreadsheet file. All internal levels have been structured into the main 5 university's activities: research, teaching, transfer, administration, and services, thereby reflecting university's diversity. Figure 3 shows the distribution of the 187 internal URLs according to the related mission, highlighting the poor representation of transfer activities (10 URLs). Both services and research represent the 65% of third-level subdomains, whereas teaching and administration are equally distributed (15% each).

**Figure 3. Distribution of internal URLs according to related university mission at Harvard University**

Both contour and internal URLs have been measured by an institutional (count page) and external (URL mention) indicator. Table VII shows the top 25 internal URLs for each indicator. Contour values are also offered to contextualize the results obtained.

Table VII. Core level at Harvard University: institutional and external measures

| INSTITUTIONAL | | EXTERNAL | |
|---|---|---|---|
| URL | COUNT PAGE | URL | URL MENTIONS |
| **harvard.edu** | **7,615,804** | **harvard.edu** | **38,470,780** |
| mcz.harvard.edu (service) | 1,920,000 | law.harvard.edu (administration) | 4,750,000 |
| seas.harvard.edu (administration) | 663,000 | fas.harvard.edu (administration) | 3,270,000 |
| lib.harvard.edu (service) | 589,000 | hcs.harvard.edu (service) | 2,470,000 |
| coursecatalog.harvard.edu (teaching) | 555,000 | **blogs.law.harvard.edu** | 1,980,000 |
| map.harvard.edu (service) | 460,000 | hsph.harvard.edu (administration) | 1,260,000 |
| fas.harvard.edu (administration) | 371,000 | cfa.harvard.edu (research) | 1,130,000 |
| catalyst.harvard.edu (research) | 325,000 | hms.harvard.edu (administration) | 1,070,000 |
| chem.harvard.edu (teaching) | 236,000 | huh.harvard.edu (service) | 914,000 |
| abcd.harvard.edu (service) | 209,000 | lib.harvard.edu (service) | 874,000 |
| law.harvard.edu (administration) | 181,000 | bidmc.harvard.edu (research) | 844,000 |
| mcb.harvard.edu (teaching) | 170,000 | mgh.harvard.edu (service) | 834,000 |
| hunap.harvard.edu (service) | 148,000 | hks.harvard.edu (administration) | **775,000** |
| oeb.harvard.edu (teaching) | 94,300 | hcl.harvard.edu (service) | 714,000 |
| dce.harvard.edu (transfer) | 93,900 | news.harvard.edu (service) | 644,000 |
| cfa.harvard.edu (research) | 88,900 | oeb.harvard.edu (teaching) | 540,000 |
| mgh.harvard.edu (service) | 86,300 | dfci.harvard.edu (research) | 539,000 |
| huh.harvard.edu (service) | 86,000 | post.harvard.edu (service) | 534,000 |
| eecs.harvard.edu (administration) | 80,400 | mcb.harvard.edu (teaching) | **531,000** |
| **hbs.edu (administration)** | 75,800 | as.harvard.edu (research) | 518,000 |
| **blogs.law.harvard.edu** | 72,900 | catalyst.harvard.edu (research) | 490,000 |
| hup.harvard.edu (service) | 71,400 | **hbs.edu (administration)** | 468,000 |
| hks.harvard.edu (administration) | 66,600 | eecs.harvard.edu (administration) | 455,000 |
| hms.harvard.edu (administration) | 65,200 | chandra.harvard.edu (research) | 454,000 |
| news.harvard.edu (service) | 56,700 | gov.harvard.edu (teaching) | 447,000 |
| chs.harvard.edu (teaching) | 52,500 | biology.harvard.edu (teaching) | 425,000 |
| hcs.harvard.edu (service) | 50,400 | chs.harvard.edu (teaching) | 350,000 |
| dfci.harvard.edu (research) | 47,500 | economics.harvard.edu (teaching) | 342,000 |

The results indicate a great percentage of count pages from service-oriented institutions and products. For example, "mcz.harvard.edu" belongs to the Museum of Comparative Zoology, which is the 25.2% of the global count gathered by the entire contour URL. The Library (3rd position), Campus map (5th), or the Public affairs & Communications service (24th) also demonstrate the importance of service products, whereas research institutions reflect general poor performances (only 3 institutions within the top 25).

Moreover, in some cases, the high content volume achieved by some institutions is due to the existence of a service within the corresponding web domain. This is the case of "law.harvard.edu" (Harvard Law School), which achieves 10th position, but the 40.27% of the count pages are due to the blog platform (although is not a third level subdomain, it is showed on 20th position in order to prove its influence).

Otherwise the summation of the 187 URLs' count is 7,467,107 pages (54.66% of the global account). Taking into account that these 187 domains are compound by the main Faculties, Schools, some departments, research institutes and centres, etc., and also considering that within third-level domains is measured the count page of the other subdomain/folds levels, this low percentage is inferred that the remaining count page is stored within first level folds ("harvard.edu/*"), and intranet services.

As regards external measures, table VII shows the Harvard Law School on the first position, again under the influence of its blog platform. Research domains reflect better performances respect to count measures (up to 6 domains within top 25). The remaining activities are equally distributed (6 teaching domains, 6 administration, and 7 services), and any transfer domain is found within top 25.

In this case, a methodological problem with "URL mention" has been identified. Whereas "harvard.edu" URL founds 38,470,780 mentions outside "harvard.edu" site, the summation of all 187 URLs achieve 36,183,780 mentions, confirming the limitations of this indicator for internal and systemic academic levels.

The correlation between count page and URL mentions is insignificant (R=0.19), confirming the need for more accurate external measures' indicators, once Yahoo! stopped link commands. Title mention instead of URL mention could be a solution to be further investigated.

Otherwise, the existence of URL duplications shows another interesting effect. On one hand, are detected examples of more than 1 valid URL for an institution or product; this is the case of "post.harvard.edu", which redirects to "alumni.harvard.edu", being both URLs within the academic website, or the "The Institute for Quantitative Social Science", which has 2 valid URLs: "iq.harvard.edu" and "cbrss.harvard.edu".

On the other hand, other redirections are found where the redirected URL is outside Harvard domain; this is the case for "Massachussets Eye and Ear Infirmary", whose "meei.harvard.edu" URL redirects to "http://www.masseyeandear.org", or "Harvard Business School", whose "hbs.harvard.edu" redirects to "hbs.edu".

The effect of the first problem (2 or more valid URLs) provokes a dispersion of performance, and all URLs should be added in count and URL mentions to obtain an aggregated and global value. The effect of the second problem (2 or more URLs, with some of them non-valid) provokes an underrepresentation of the university. For example, "hbs.harvard.edu" obtains 3 count results and 161,000 URL mentions, whereas "hbs.edu" obtains 75,800 count results (which do not count for "harvard.edu" performance), and 468,000 URL mentions. Table VII shows the position that "hbs.edu" should have been obtained if it had been considered within "harvard.edu" web domain.

4.3.2. Satellite level at Harvard University

As regards satellite level, table VIII shows (for each considered platform) institutional and external measures at contour level. Academia provides by far both the largest count page and URL mention values (21,000 and 116,000 respectively). Otherwise, it is identified a low performance on Youtube respect to the count pages (only 1, which contrast with its external activity), and Flickr respect to URL mentions (only 10, which contrast with its institutional activity). The purposes of these platforms and the URL syntaxes could explain this effect.

**Table VIII. Satellite level at Harvard University: institutional and external measures**

| LEVEL | URL | COUNT PAGE | URL MENTIONS |
|---|---|---|---|
| ACADEMIA | harvard.academia.edu | 21,000 | 116,000 |
| | Summation internal URLs (378) | **517** | **81** |
| FACEBOOK | facebook.com/Harvard | 5,580 | 9,560 |
| | Summation internal URLs (56) | **26,840** | **12,656** |
| FLICKR | flickr.com/groups/harvard | 951 | 10 |
| | Summation internal URLs (15) | **7,209** | **511** |
| TWITTER | twitter.com/Harvard | 8,240 | 7,690 |
| | Summation internal URLs (110) | **155,996** | **30,991** |
| YOUTUBE | youtube.com/harvard | 0 | 5,670 |
| | youtube.com/user/harvard | 1 | 15,300 |
| | Summation internal URLs (18x2) | **20** | **15,886** |

With respect to internal level, a total of 597 URLs have been gathered. Academia becomes the platform with more internal units (378), followed by Twitter (110), and Facebook (56). Table VIII also shows the summation of count and URL mentions for each internal URL per platform. Twitter and Facebook show the better performances, followed by Youtube. Should be highlighted the low performance of Academia, probably due to the larger URL syntaxes, and the low participation of users within the platform by means of uploading research material and other contents.

In any case, the overall results confirm both contour and internal satellite levels add significant amount of count pages and URL mentions, which reflect an impact of universities outside the official academic website.

## 5. Conclusions

The proposed analysis model, based on parts (core and satellite), levels (institutional and external) and sublevels (contour and internal), is simple, independent of technique (allows constant updating of the model without having to modify its general architecture), and provides structured information which enables a complete analysis of each institution.

Cybermetrics (insofar as it is a technique through which the conceptual analysis model may be applied) facilitates the measurement of the same indicators at different levels of institutional analysis (i.e. it enables systemic analyses), something not possible with other quantitative techniques such as bibliometrics, which depend on the assignation of the institution on the part of the authors of the scientific contributions.

Furthermore, the localisation of lower units favours the identification of institutions with missions and objectives that are not only scientific, and this may assist in the process of profiling certain traits related to university diversity and describing the general performance of the university with greater precision.

Notwithstanding, the study of university web units (internal level) is complex. The analysis of Complutense University of Madrid shows an excessive casuistic of URL syntaxes and redirections. This reflects in a lower web performance, mainly due to a lack of appropriate web management policies. That is, the internal analysis can be used not only to measure web production or impact but also to detect the degree of maturity of the university structure on the Web (low in the case of Complutense University).

The complete analysis of Harvard University has proven the advantages of the model. The core internal level shows as some subdomains related to service activities achieve great amount of performance. The model allows the identification of unequal distribution of impact (both institutional and external) regarding the universities' missions. Satellite level shows that the activities done by universities outside the official website are significant and measurable. For all these reasons, could be asserted that "harvard.edu" is not enough to analyse "Harvard University" under a webometric point of view, and that the proposed model helps to expand and complete this analysis by identifying and describing the structure of the website (that reflects the structure of the organization), and assessing the web performance of each element identified.

Despite the model's advantages, it also has certain limitations, not conceptual but practical:

- The model has too many instrumental limitations: the shortcomings in the construction of the online seats, the inaccuracy of some indicators (URL mentions for external impact), and the inaccessibility of certain indicators (e.g. web traffic data) mean that some indicators or units cannot be adequately measured.
- An excessive diversity and randomness has been detected in the URL syntax of the internal unit seats (the aforementioned example of Complutense University of Madrid can be logically extended to the general university web space); this should be corrected with a suitable institutional web policy.
- All the levels can contain multidomains and URL redirections (e.g. "hbs.harvard.edu" and "hbs.edu"), which make the identification of universities and quantification of content more complicated.
- The internal institutional level of the model is only applicable to a specific university system, in which the university structure is clearly specified (legally or not), and does not have to coincide with the university systems of other countries.
- Cybermetric method does not permit the direct capture of some diversity categories, such as those of components or programming. Other categories and attributes (such as climate diversity, thematic areas, etc.) must be manually and independently considered for each university and unit.

The next steps to be taken in this research project are the thorough and comprehensive application of the model into a complete university system, and the analysis of the aforementioned limitations, with the aim of determining its scope and applicability for wider cybermetric analysis.

## 6. Notes

[1] These products are not selected according to any criteria of quality or relevance, the selection is merely illustrative.

[2] The Research Results Transfer Offices (OTRI) are aimed to boost relations between the university scientific community, business and other socio-economic actors to take advantage of R&D capabilities and results of research activity in universities.

[3] http://www.youtube.com/user/StanfordUniversity (accessed 1 April 2012).

[4] http://vimeo.com/cambridge (accessed 1 April 2012).

[5] http://www.flickr.com/groups/harvard (accessed 1 April 2012).

[6] http://www.slideshare.net/norwichuniversity (accessed 1 April 2012).

[7] http://www.facebook.com/columbia (accessed 1 April 2012).

[8] http://mit.academia.edu (accessed 1 April 2012).

[9] http://www.usal.es/~oca (accessed 1 April 2012).

[10] Currently the website is not accessible, and cannot be located the reference year.

[11] http://www.webometrics.info (accessed 1 April 2012).

[12] http://www.4icu.org/top200 (accessed 1 April 2012).

[13] http://www.universidad.edu.co (accessed 3 December 2008).

[14] http://vcmike.blogspot.com/2006/01/ranking-colleges-using-google-and-oss.html (accessed 1 April 2012).

[15] http://www.universitymetrics.com/g-factor (discontinued; accessed 1 April 2012)

[16] http://googlecollegerankings.com (accessed 1 April 2012).

[17] http://www.um.es/gtiweb/adrico/#Datos (accessed 1 April 2012).

[18] http://en.wikipedia.org/wiki/Generic_top-level_domain (accessed 1 April 2012).

[19] http://en.wikipedia.org/wiki/Country_code_top-level_domain (accessed 1 April 2012).

[20] http://www.ietf.org/rfc/rfc3986.txt (accessed 1 April 2012).

[21] Complutense University of Madrid is considered only as example.